\documentclass[useAMS,usenatbib]{mn2e}

\usepackage{graphicx}

\bibpunct{(}{)}{;}{a}{}{ }
\begin{document}
\title[NGC253 ULXs: variability and extended coronae]{In-depth studies  of the NGC 253 ULXs with XMM-Newton: remarkable variability in ULX1, and evidence for  extended coronae }
\author[R. Barnard]{R. Barnard$^1$\\
$^{1}$Department of Physics and Astronomy, The Open University, Walton Hall, Milton Keynes, MK7 6AA, UK}
\date{}

\pagerange{\pageref{firstpage}--\pageref{lastpage}} \pubyear{2009}

\maketitle

\label{firstpage}

\begin{abstract}
We examined the variability of three ultra-luminous X-ray sources (ULXs) in the 2003, 110 ks XMM-Newton observation of NGC\,253.   Remarkably, we discovered ULX1 to be three times more variable than ULX2 in the 0.3--10 keV band, even though ULX2 is brighter. Indeed, ULX1 exhibits a power density spectrum that is consistent with the canonical high state or very high/steep power law states, but not the canonical low state. The 0.3--10 keV emission of ULX1 is predominantly non-thermal, and may be related to the very high state.  We also fitted  the ULX spectra with disc blackbody, slim disc  and  convolution Comptonization  ({\sc simpl$\otimes$diskbb}) models. The brightest ULX spectra are usually described by a two emission components (disc blackbody + Comptonized component); however, the {\sc simpl} model results in a single emission component, and may help determine whether the well known soft excess is a feature of ULX spectra or an artifact of the two-component model.  The {\sc simpl}  models were rejected for ULX3 (and also  for the black hole + Wolf-Rayet binary IC10 X-1); hence, we infer that the observed soft-excesses are genuine features of ULX emission spectra. We use an extended corona scenario to explain the soft excess seen in all the highest quality ULX spectra, and provide a mechanism for stellar mass black holes to exhibit super-Eddington luminosities while remaining locally sub-Eddington.
\end{abstract}


\section{Introduction}

In \citet{bck08} we identified a new spectral state in the confirmed black hole + Wolf-Rayet binary IC10 X-1, as well as the BH+WR candidate NGC\,300 X-1 and the well known black hole binary LMC X-1. Both IC10 X-1 and NGC300 X-1 exhibited $\sim$ 0.0001--0.1 Hz power density spectrum (PDS) where the power is inversely proportional to the frequency, like the canonical high state. However, the  0.3--10 keV  for these sources emission is predominantly non-thermal, while the canonical emission spectrum for high state black hole X-ray binaries is dominated by a multi-temperature disk blackbody that contributes $\sim$90\% of the 1--10 keV flux \citep{mr03}.

In \citet{bck08} we proposed that the non-thermal high state was due to a persistently high accretion rate maintaining a stable corona; most black hole X-ray binaries are transient, and we suggested that the  canonical  soft high state was caused by the ejection of the corona during outburst. After finding out from \citet{sk08} that $\sim$90\% of X-ray sources with luminosities $>$10$^{39}$ erg s$^{-1}$ have non-thermal spectra, we suggested that the same physical processes may apply in these sources  as in IC10 X-1 and  NGC\,300 X-1. The ultra-luminous X-ray sources (ULXs) are an intriguing class of extragalactic X-ray sources that are not associated with the galaxy nucleus, yet exhibit X-ray luminosities $>$2$\times$10$^{39}$ erg s${-1}$, the Eddington limit for a $\sim$15 M$_{\odot}$ black hole \citep[see e.g.][for a review]{rob07}.

Our survey of X-ray sources in the 2003 XMM-Newton observation of the nearby spiral galaxy  NGC\,253 \citep{bsgk08} yielded three X-ray sources with 0.3--10 keV luminosities $>$2$\times$10$^{39}$ erg s$^{-1}$. We shall call these ULX1--ULX3, and their known properties are presented in Table~\ref{props}; these properties are taken from our survey \citep{bsgk08}. In this paper, we examine the variability of these sources. We also perform additional spectral fitting; we model their spectra with slim disc models, and ``physical'' Comptonization models where the inner disc provides the seed photons.
In Section 2 we discuss the observation and data analysis, followed by the results in Section 3. Our discussion and conclusion is presented in Section 4.

\subsection{ULX variability}

Few ultra-luminous X-ray sources (ULXs) have sufficiently high quality  data for timing analysis. The  PDS  of M82 X-1 has been characterised by a quasi-periodic oscillations (QPOs) at 54 mHz \citep{sm03} and 114 Hz \citep{dt06}, on top of a broken power law continuum \citep{dt06}. QPOs have also  been found in Holmberg IX X-1 \citep[along with a power law continuum][]{dgr06} and NGC\, 5408 X-1 \citep[with a broken power law continuum][]{stro07}.

\citet{hvr09} conducted a timing study of ULXs observed with XMM-Newton.
 They analysed 19 observations of 16 ULXs and found  6 ULXs with PDS best described by power laws or broken power laws: M82 X-1, NGC\, 1313X-1, NGC\,1313 X-2, NGC\,55 ULX, Ho IX X-1 and NGC\,5408 X-1; the others showed no significant variability. \citet{hvr09} obtained upper limits to the variability of these ULXs, and concluded that the intrinsic variability was lower.

\subsection{ULX emission spectra}

Two emission components are often used to describe the X-ray spectra of ULXs: a thermal component, and a component, e.g. power law,  to represent inverse Compton scattering of cool photons on hot electrons. \citep[see e.g.][]{stob06,grd09}. These fits often result in a ``soft excess'', where the power law component diverges from the thermal component  Such spectra have been deemed by some as non-standard and unphysical, assuming that the seed photons are provided by the inner disc  \citep[see e.g.][]{rww05,gs06}.

\citet{grd09} conducted a survey of XMM-Newton observations of ULXs with $>$10000 source counts in their EPIC spectra; their preferred model consists of a disc blackbody and a Comptonized component  with the seed photons tied to the inner disc temperature. They found two universal features: a soft excess and roll-over in the spectrum above $\sim$3 keV. They found that disc + Comptonized corona emission models fitted the data well; however, the resulting coronae were cool (k$T$ $\simeq$ 1--3 keV) and optically thick ($\tau$ $\sim$ 5--30). The combination of a cool disc and broken hard component led \citet{grd09} to define a new, ultraluminous spectral state. Such coronae are distinct from coronae of Galactic binaries ($\tau$ $\sim$ 1), and \citet{grd09} use this difference to reject the scenario of sub-Eddington accretion onto intermediate mass black holes.

\begin{figure}
\resizebox{\hsize}{!}{\includegraphics[angle=0,scale=0.6]{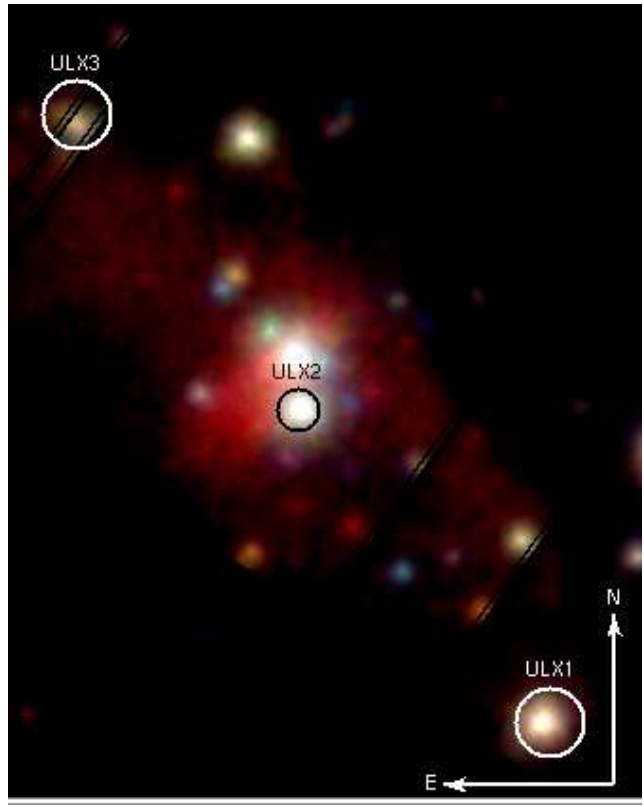}}
\caption{ Detail of a three colour pn + MOS image of NGC\,253 showing ULX1, ULX2 and ULX3. Red represents 0.3--2.0 keV, green represents 2.0--3.0 keV, and blue represents 4.0--10 keV. The image is log-scaled, and the source extraction regions for the ULXs are labeled. }\label{pic}
\end{figure}

\section{Observation and analysis}

In \citet{bsgk08} we conducted a spectral survey of X-ray sources in the 2003 XMM-Newton observation of NGC\, 253; full details of our data reduction are given in that paper. The present  work concerns new variability studies of three ULXs. In \citet{bs07} we showed that the combination of lightcurves from the different cameras on-board XMM-Newton can lead to artefacts in the PDS. To ensure fidelity of the PDS,  we performed our variability analysis on the  unbinned  pn  events files for each source. The total exposure time was $\sim$110 ks; however, several background flares occurred during the observation; the flares were removed \citep[see ][for details]{bsgk08},  leaving $\sim$50 ks of good data.
\begin{table*}
 \centering
% \begin{minipage}{140mm}
  \caption{ Known properties of the three ULXs, taken from our survey of X-ray sources in NGC\,253 \citep{bsgk08}. For each source we give the source coordinates as returned by the source detection routine; these have uncertainties of $\sim$2$''$. We then give the number of net source counts in pn and MOS detectors. Next we give the details of the best fit spectral model: absorption ($N_{\rm H}$), blackbody temperature (k$T$) and photon index ($\Gamma$), along with the corresponding $\chi^2$/dof and 0.3--10 keV luminosity / 10$^{39}$ erg s$^{-1}$. Numbers in parentheses indicate 90\% confidence limits on the final digit. }\label{props}
  \begin{tabular}{ccccccccc}
  \noalign{\smallskip}
  \hline
  \noalign{\smallskip}
Source & Position & pn & MOS &  $N_{\rm H}$/10$^{21}$ atom cm$^{-2}$   & k$T$/keV &  $\Gamma$ & $\chi^2$/dof & $L$/10$^{39}$\\
\noalign{\smallskip}
 \hline
\noalign{\smallskip}
ULX1 &  00 47 22.56 $-$25 20 51 & 10537& 10390 & 2.00(2)& 0.73(5)& 2.14(4)& 347/323& 2.90(12)\\
ULX2 & 00 47 32.98 $-$25 17 50 & 11614 & 12014 &2.9(2) & 0.98(6) & 1.94(5) & 374/374 & 4.10(19) \\
ULX3 & 00 47 42.41 $-$25 14 59 & 773 & 4156 &6.0(11) & 0.94(8) & 3.4(5) & 84/76 & 2.4(4)\\
\noalign{\smallskip}
\hline
\noalign{\smallskip}
\end{tabular}
%\end{minipage}
\end{table*}

\section{Results}

  Figure~\ref{pic} shows a three colour detail of the combined EPIC (pn + MOS1 + MOS2) image of NGC\,253, overlaid with the source extraction regions; the image is log-scaled, north is up and east is left.

\subsection{Variability}
\begin{figure}
\resizebox{\hsize}{!}{\includegraphics[angle=270,scale=0.6]{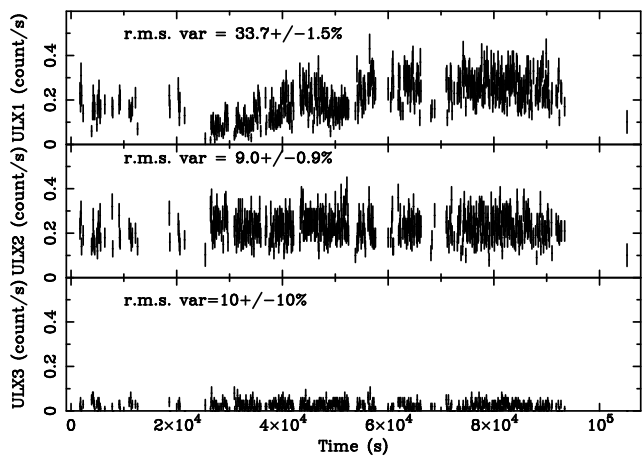}}
\caption{ Lightcurves of the three ULXs, binned to 100 s from the unbinned 0.3--10 keV  pn events. X- and y- axes are set to the same scales. We give the fractional r.m.s. variability of each lightcurve. }\label{lcs}
\end{figure}

We present the 0.3--10 keV pn lightcurves of ULX1, ULX2 and ULX3 in Figure~\ref{lcs}, binned to 100 s from the unbinned events. Intervals with strong background flares have been filtered out. The background contribution is negligible in each case. The x- and y-axes are plotted on the same scale for each source. For each lightcurve we provide the fractional r.m.s. variability calculated by the program {\sc lcurve} from the FTOOLs analysis suite\footnote[1]{http://heasarc.nasa.gov/lheasoft/ftools/ftools\_menu.html}. We note that  ULX3  is rather fainter than ULX1 and ULX2 and there are two reasons for this. Firstly, ULX3 is located at the corner of a chip in the pn image, and  the collecting area is reduced when the (FLAG==0) filter is applied;  lastly,  ULX3 is more heavily absorbed than the other two. ULX1 is particularly variable, ranging in intensity over $\sim$0.0--0.4 count s$^{-1}$; analysis of the ULX1 lightcurves in the 0.3--2.5 keV and 2.5--10 keV bands showed no evidence for energy dependence in this variability.

\begin{figure}
\resizebox{\hsize}{!}{\includegraphics[angle=270,scale=0.6]{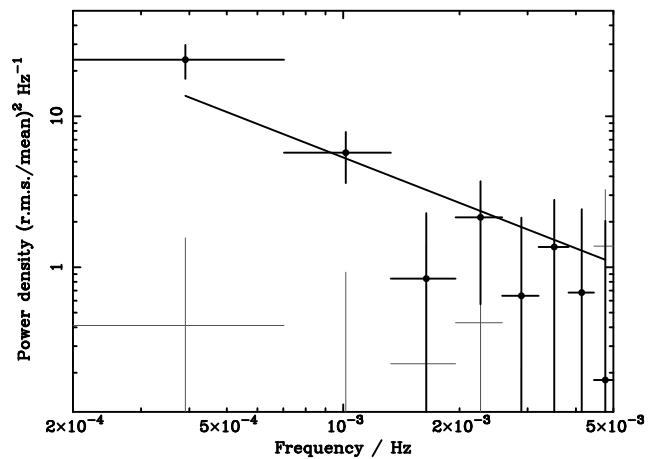}}
\caption{ Power density spectra constructed from 0.3--10 keV pn events for ULX1 (black) and ULX2 (grey); the expected noise is subtracted. The PDS of ULX1 features strong variability at low frequencies; we provide the best fit power law model with $\gamma$ = 1.0$\pm$0.5 ($\chi^2$/dof = 3/6).   }\label{pds}
\end{figure}

For each source, we constructed 0.3--10 keV PDS that were averaged over several intervals consisting of 64 bins with a resolution of 100 s; intervals of background flaring were excluded. Figure~\ref{pds} shows the resulting PDS for ULX1 (black) and ULX2 (grey) , which have comparable intensities. The PDS are  normalized to give the (r.m.s./mean)$^2$ power and the expected noise is subtracted; the PDS are binned by to give 56 frequencies per bin. The PDS of ULX1 (black) is well described by a power law; if the power at frequency $\nu$ is $P(\nu)$, then $P(\nu) \propto \nu^{-\gamma}$. We show the best fit power law to the ULX1 PDS in Fig.~\ref{pds}; $\gamma$ = 1.0$\pm$0.5, with $\chi^2$/ = 3 for 6 degrees of freedom (dof). Fitting the ULX1 PDS with zero power yields $\chi^2/$dof = 26/8, an unacceptable fit; hence the observed power in ULX1 is significant. The PDS for ULX2 shows no strong evidence for variability, even though ULX2 is slightly brighter than ULX1; fitting zero power to the ULX2 PDS yields $\chi^2$/dof = 5/6. Hence the observed power from ULX1 is intrinsic, rather than an artefact of the observation. Indeed, the  PDS of the combined pn+MOS1+MOS2 lightcurves of ULX2 and ULX3 were also featureless. These results show that 10,000 source counts are sufficient for detecting $\gamma$$\sim$1  PDS in sources where the r.m.s. variability is $\sim$30\%, but not for sources with 10\% variability.

The frequency range of our PDS is $\sim$2$\times$10$^{-4}$--5$\times$10$^{-3}$ Hz. \citet{cgr01} showed that Cygnus X-1 exhibits similar power law noise with $\gamma$ $\sim$1 in this frequency range during its low state and high state; however, the r.m.s.$^2$  power  of Cygnus X-1 at $\sim$10$^{-4}$ Hz is $\sim$20 times higher in the high state than in the low state. This is unsurprising, because  the high state PDS is characterisd by a $\gamma$ $\sim$1 power law over the 0.0001--10 Hz range, while its low state is similar, except for the interval $\sim$2$\times$10$^{-3}$--2$\times$10$^{-1}$ Hz where $\gamma$ $\sim$0. The variability of ULX1 is consistent with that of Cygnus X-1 in the high state, but not its low state.

 We compared  the 10$^{-2}$--10 Hz PDS of Cygnus X-1 in the high and low states \citep{cgr01}, with the  10$^{-2}$--10 Hz PDS of other black hole X-ray binaries in the high, low and steep power law states \citep{mr03}. The high state PDS of Cygnus X-1 has rather more power at 10$^{-2}$ Hz than the high states of other black hole binaries; this may be a consequence of its high mass donor. The low state PDS of Cygnus X-1 shows   variability at 10$^{-2}$ Hz that is as high as, or higher than the PDS of other black holes in the low state. It is difficult to distinguish between the high and very high states variability alone; we therefore infer that ULX1 is in a high accretion rate state that may be related to the high or very high/ state.

\subsection{Spectral analysis}

\begin{figure}
\resizebox{\hsize}{!}{\includegraphics[angle=270,scale=0.6]{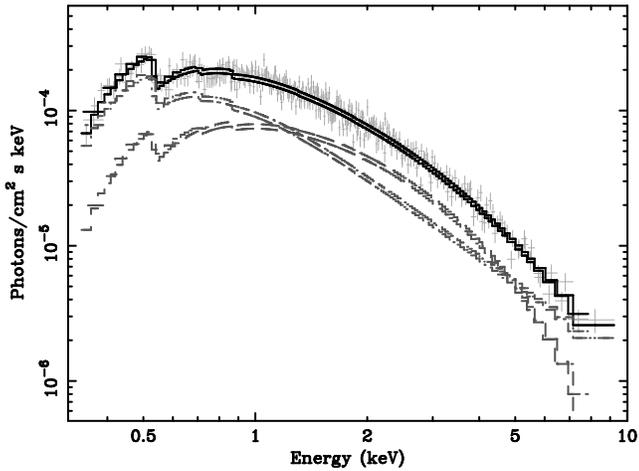}}
\caption{ Unfolded  0.3--10 keV pn and combined MOS spectra of ULX1,  modelled with  disc blackbody and  power law emission components, and suffering photo-electric absorption by material in the line of sight. A constant of normalization accounts for the differences in pn and MOS responses.  }\label{ufspec}
\end{figure}

The emission spectra of the brightest  ULXs are generally well described by disc emission plus a Comptonization component \citep[see e.g.][]{stob06}.  Figure~\ref{ufspec} shows the unfolded 0.3--10 keV  pn and  spectrum of ULX1, simultaneously modeled with a  disc blackbody and a power law component to represent Comptonization; line-of-sight absorption is included, and a constant of normalization accounts for differences in the pn and MOS responses. The Comptonized component contributes most of the 0.3--10 keV emission; however, the thermal emission dominates the $\sim$1--5 keV range. The power law component dominates the spectrum below $\sim$ 1 keV; this "soft excess" is sometimes deemed unphysical, as the seed photons for the Comptonization are assumed to come from the inner disc \citep[see e.g.][]{rww05,gs06}.
In this section, we model the spectra of ULX1--ULX3 with several alternative models.

\subsubsection{Disc blackbody models}
The emission spectra of distant ULXs are often successfully described by disc blackbody models \citep[see e.g.][]{imt09}. However, for these models to be meaningful, they must have inner disc temperatures and luminosities that are consistent with the black hole mass, $M_{\rm BH}$; the temperature is governed by $M_{\rm BH}^{-1/4}$, while the luminosity should not exceed $\sim$0.5$L_{\rm Edd}$ \citep{mit84}.

Successful fits to the NGC253 ULX spectra with meaningful disc blackbody models would indicate that these systems were in the canonical high state; disc temperatures range over $\sim$0.7--1.5 keV for most known black hole binaries in the high state \citep{mr03}.  Therefore we simultaneously fitted  absorbed disc blackbody models to the pn and combined MOS spectra for each ULX; a constant of normalisation is included to account for differences in the pn and MOS responses.
We reject the absorbed disc blackbody model for ULX1 and ULX3. The disc blackbody fit for ULX2 is acceptable: $N_{\rm H}$ = 1.56$\pm$0.08$\times 10^{21}$ atom cm$^{-2}$,  k$T_{\rm in}$ = 1.68$\pm$0.04 keV, and $\chi^2$/dof = 397/376 (good fit probability, g.f.p.  = 0.22). However, the temperature for ULX2 is  rather higher than known black hole binaries, and is inconsistent with the modelled 0.3--10 keV luminosity of 3$\times$10$^{39}$ erg s$^{-1}$.  F-testing shows the blackbody + power law model to be a better fit; the probability that this improvement is due to chance is 1.4$\times$10$^{-5}$.

%\begin{table*}
% \centering
% \begin{minipage}{140mm}
%  \caption{ Best  disc blackbody fits to the pn and MOS spectra of the three ULXs. We show the ULX, absorption, temperature, constant of normalization and $\chi^2$/dof; the good fit probability is presented in square brackets. Uncertainties are provided when the fit is acceptable}\label{specmod}
 % \begin{tabular}{cccccc}
 % \noalign{\smallskip}
 % \hline
 % \noalign{\smallskip}
 % Model & $N_{\rm H}$/10$^{20}$ atom cm$^{-2}$   & k$T_{\rm in}$/keV &  $n_{\rm MOS}$& $\chi^2$/dof\\
%\noalign{\smallskip}
% \hline
%\noalign{\smallskip}
%ULX1 &  5 & 1.2 & 1.08 & 422/325 [2$\times$10$^{-4}$]\\
%ULX2 & 15.6(8) & 1.68(4) & 1.06(2) & 397/376 [0.22]\\
%ULX3 & 19 & 1.14 & 1.40& 135/78 [6$\times$10$^{-5}$]\\
%\noalign{\smallskip}
%\hline
%\noalign{\smallskip}
%\end{tabular}
%\end{minipage}
%\end{table*}

\subsubsection{Slim disc models}
Slim disc models have also been applied to some ULXs \citep[see e.g.][]{oek06}. \citet{kaw03} provides a suite of tabular models for XSPEC\footnote{http://heasarc.gsfc.nasa.gov/docs/xanadu/xspec/models/slimdisk.html}, including a standard disc, thermal and Comptonized slim discs, with and without relativistic effects. Each model is described by four parameters; the mass of the accretor, $M$, is given in solar masses; the accretion rate, $\dot{M}$, is normalised to $L_{\rm Edd}$/c$^2$, so that Eddington accretion corresponds to $\dot{M}$ $\simeq$ 10 \citep{kaw03}; the viscosity parameter, $\alpha$, is limited to the range 0.01--1; the normalisation is defined as (10 kpc/$d$)$^2$, where $d$ is the distance. The distance was fixed to 4 Mpc, following \citet{grimm03}.

We applied these slim disc models to our ULX targets. For ULX1, no thermal model resulted in good fits, meaning that Comptonization was required. The best results were gleaned from the Comptonized slim disc model  with relativistic correction: $N_{\rm H}$ = 1.53$\pm$0.06$\times 10^{21}$ atom cm$^{-2}$, $M$ = 17.5$\pm$1.5 M$_{\odot}$, $\dot{M}$ = 12.8$\pm$0.8 $L_{\rm Edd}/c^2$, $\alpha$ = 0.11$\pm$0.03, and $\chi^2$/dof = 351/324 (g.f.p. = 0.14). For ULX2, all slim disc models failed, with g.f.p. $<$2$\times$10$^{-6}$.   ULX3 was best described by a modified blackbody with relativistic correction: $N_{\rm H}$ = 3.7$\pm$0.2$\times 10^{21}$ atom cm$^{-2}$, $M$ = 8$\pm$7 M$_{\odot}$, $\dot{M}$ = 10.7$\pm$0.7 $L_{\rm Edd}/c^2$, $\alpha$ = 0.36$\pm$0.10, $\chi^2$/dof = 94/77 (g.f.p. = 0.09). F-testing showed that our two-component model has a 99.6\% probability of providing a significant improvement.

\subsubsection{Self-consistent Comptonization of a disc blackbody}

 \citet{snm08} have produced a convolution model for XSPEC v12, {\sc simpl}, which describes the Comptonization of any input spectrum; the modeling can include up-scattering and down-scattering, or upscattering only. It uses only two free parameters: the photon index of the resulting power law, and the fraction of scattered photons. Fitting our ULXs with {\sc simpl$\otimes$diskbb} XSPEC emission model will help determine whether the observed soft excess is a feature of the ULX spectrum or the model.

\begin{figure}
\resizebox{\hsize}{!}{\includegraphics[angle=270,scale=0.6]{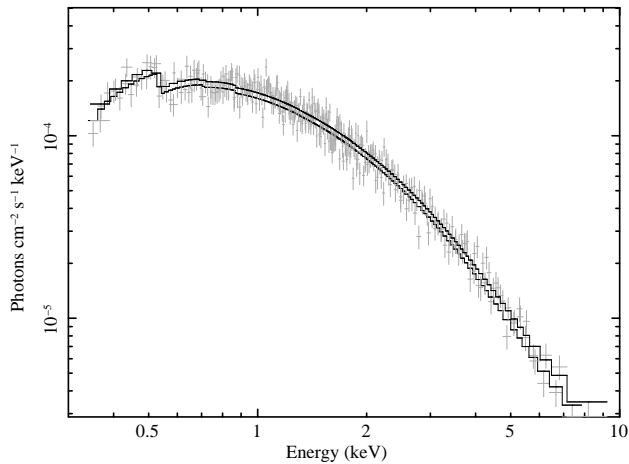}}
\caption{ Unfolded  0.3--10 keV pn and combined MOS spectra of ULX1  with the best-fit {\sc simpl} model. A constant of normalization accounts for the differences in pn and MOS responses.  }\label{simpl}
\end{figure}

Best fit {\sc simpl} models are provided in Table~\ref{simpltab}; these include up-scattering and down-scattering. We reject all {\sc simpl} fits to ULX3. We found acceptable fits for ULX1 and ULX2; however, the emission parameters could not be constrained.  The best fit to ULX1 is presented in Fig.~\ref{simpl}. F-testing tells us that we must prefer the two-component models, since they provided a lower $\chi^2$ for the same degrees of freedom in each case, and this is  always a significant improvement. We therefore  infer that the soft excess is a genuine feature of ULX spectra; such a feature could be produced by an extended corona that has access to low energy photons from the outer disc.

One might expect the {\sc simpl} and two component models to give equally acceptable results, as they are both describing the Comptonization of a disc blackbody, with no specified corona geometry. However, the {\sc simpl} model does not include photons outside the energy range of the spectrum; hence, the softer photons in the cooler regions of the disc are excluded.  We further note that the {\sc simpl} model assumes that all photons have equal probability of being scattered, and that photons of all energies are scattered by the same amount \citep{snm08}.  The latter may approximate a compact corona, but is unlikely to be true for an extended corona, as the photon energies will vary by several orders of magnitude.

We also fitted {\sc simpl} models to  our XMM-Newton spectra for the known black hole + Wolf-Rayet binary  IC10 X-1 \citep[see][ and references within]{bck08}. The best fit model yielded $\chi^2$/dof = 763/650 (g.f.p. = 0.0014) and is rejected; the two component model yields $\chi^2$/dof = 699/650 (g.f.p = 0.08). Hence we infer the soft excess to be a genuine feature of the IC10 X-1 emission spectrum also.

\begin{table*}
 \centering
% \begin{minipage}{140mm}
  \caption{ Best fit self-consistent Comptonization models to the ULXs. The spectral model used was {\sc const*wabs*simpl(diskbb)} with {\sc const} accounting for the differences in pn and MOS response. The models include up-scattering and down-scattering. We show the absorption ($N_{\rm H}$/10$^{20}$ atom cm$^{-2}$), photon index ($\Gamma$), scattering fraction ($f_{\rm scat}$), and inner disc temperature (k$T_{\rm in}$). We then give the $\chi^2$/dof and good fit probability. The emission parameters are unconstrained for all models. }\label{simpltab}
 \begin{tabular}{ccccccccc}
  \noalign{\smallskip}
  \hline
  \noalign{\smallskip}
  Model& $N_{\rm H}$/10$^{20}$ atom cm$^{-2}$   & $\Gamma$ & $f_{\rm scat}$ & k$T_{\rm in}$&   $\chi^2$/dof [g.f.p]\\
\noalign{\smallskip}
 \hline
\noalign{\smallskip}
ULX1 & 6 & 1.1 & 0.3 & 1.0 & 359/323 [0.08]\\
ULX2 & 16 & 3.2 & 0.6 & 1.3 & 386/374 [0.32] \\
ULX3 & 24 & 1.1 & 0.4 & 0.8 & 110/76 [7E-3]\\
\noalign{\smallskip}
\hline
\noalign{\smallskip}
\end{tabular}
%\end{minipage}
\end{table*}

%\subsubsection{Is our interpretation for ULX1 feasible?}

\section{Discussion and conclusions}

ULX1 is a factor of $\sim$3 times more variable than ULX2, despite the fact that ULX2 is brighter. Furthermore, the $\sim$0.0002--0.005 Hz PDS of ULX1 is well described by P($\nu$) $\propto$ $\nu^{-1.0\pm 0.5}$; the power at 0.0002 Hz is a factor $\sim$10 higher than expected for the canonical low state  for known black hole  binaries \citep{mr03}. However, the 0.0002 Hz power of ULX1 is consistent with the high state of Cygnus X-1 \citep{cgr01}. It is difficult to differentiate between the high state and very high state (a.k.a. steep power law state) using only the PDS \citep{mr03}; however, the emission of ULX1 is more like the very high / steep power law state than the high state.

The best studied ULX spectra exhibit are well described by  thermal + Comptonization emission models. They often exhibit soft excesses which have been labeled unphysical, assuming that the seed photons are provided by the inner disc \citep[e.g.]{rww05,gs06}. We have tested this assumption using {\sc simpl} models. We have  rejected these models for ULX3 and IC10 X-1, and find that our two-component models are preferred for ULX1 and ULX2. We therefore infer that these soft excesses are genuine features of ULX spectra.

The disc blackbody contributes 1.3$\pm$0.2$\times$10$^{39}$ erg s$^{-1}$ to the unabsorbed 0.3--10 keV flux from our best two-component fit to ULX1; a distance of 4 Mpc is assumed, but the distance is not well known. This is plausible for  a black hole with mass $\sim$20 M$_{\odot}$; the most massive known stellar mass black hole, in IC10 X-1, has a mass of $\sim$20--35 M$_{\odot}$ \citep{sf08}.

% \citet{gs06} found that they could fit ULX spectra by adding or subtracting disc blackbodies from power law emission; they  suggested that the thermal component was unrelated to the disc, but was instead a convenient paramaterisation of bumps in the energy spectrum. \citet{stob06} applied various models to XMM-Newton spectra for 13 of the brightest  ULXS; their most successful model consists of a cool  disk blackbody and Comptonization in an optically thick corona({\sc diskpn +eqpair}) and fitted 11/13 ULXS; they interpret the low disc temperatures ($<$0.3 keV) as evidence for intermediate mass black hole (IMBH) accretors. However, \citet{grd09} find that the optical depths of the coronae in these models are too thick for their IMBH  interpretation  to be generally valid. They also argue that the measured temperatures are not representative of the disc temperature, as the corona obscures the inner disc; their calculated  disc temperatures indicate stellar mass black holes in most cases, and a soft excess is required for these hot discs. For those sources with cool discs, they define a new ultraluminous state, characterised by a cool disc and a broken power law within the XMM-Newton pass band, cautioning that at least 10000 counts are required to detect the break.

We propose an explanation for the soft excess that crucially involves an extended corona.
In \citet{bck08} we proposed that the canonical high soft state observed in the transient black hole binary systems is caused by the ejection of the disc corona as the X-ray luminosity increases by several orders of magnitude during  outburst. We also proposed that a steady high accretion rate could maintain a stable corona, resulting in a non-thermal high state. This theory is supported by observations of the high mass black hole binary LMC X-1, which appears to be in a stable, non-thermal high state \citep[ and references within]{bck08}.

The "non-standard" fits to the ULXs resemble the emission generally seen in the Galactic X-ray binaries with neutron star primaries; \citet{wsp88} found neutron star LMXB spectra to be dominated by a Comptonized emission component,  with an additional blackbody component  at high luminosities, while \citet{cbc01} found that the blackbody component was present at low luminosities also. The thermal and non-thermal emission components are spatially distinct in the neutron star LMXBs, as demonstrated by the high inclination LMXBs known as the dipping sources where the X-ray source is experiences increased absorption by material in the outer accretion disc on the orbital period; the thermal component is completely  and abruptly removed by dipping, while the Comptonized component experiences more gradual changes in absorption \citep[see e.g.][]{ws82,cbc95,bbc01}.

Corona sizes can be estimated for high inclination X-ray binaries,   from the ingress times of the intensity dips caused by photo-electric absorption of X-rays by cold material in the outer accretion disc \citep[see e.g.][for details]{mjc01}. Measured corona diameters for the dipping sources range from 20,000 km to 700,000 km with radius increasing with luminosity \citep{mjc01,cbc04}; the measured coronae extend over 10--50\% of the accretion disc radius, with this fraction increasing with luminosity. Such extended coronae would have access to an enormous reservoir of photons $\ll$1 keV that could account for the soft excess.

We note that further, independent evidence for extended coronae has been discovered by \citet{shn09}. They found broadened emission lines in Chandra observations of Cygnus X-2, requiring Doppler velocities of $\sim$1100--2700 km s$^{-1}$. \citet{shn09} infer from these results that the corona of Cygnus X-2 is hot, dense and extended up to $\sim$10$^{10}$ cm.

 If the coronae in ULXs are similarly extended, as our results suggest, then the Eddington limit may be relaxed and many observed ULXs may be explained by stellar mass black holes accreting at a consistently high rate, but radiating at a locally sub-Eddington level. We propose that the non-thermal component of the observed  ULX spectrum is provided by an extended corona that Comptonizes photons from the inner regions of the disc: the hot photons from the inner disc and also  cool photons from further out. The observed thermal component would then be the portion of the disc emission that is unscattered.

 This state may well be related to the canonical very high / steep power law state.  Indeed, if the high soft state is caused by ejection of the corona during transient outburst, then the very high state could indicate that the corona was regenerating; an extended corona could make the very high state appear brighter in X-rays than the high soft state if sufficient cool photons were inverse-Compton scattered into the observed energy range. It is unfortunate that the highest quality X-ray observations of Galactic black hole binaries,  from RXTE, are insensitive to energies below 2 keV; sensitive spectra could reveal a soft excess, and extended corona, in the very high state also.

 The extended corona scenario means that many ULXs could contain stellar mass black holes that exhibit a super-Eddington luminosity while being locally sub-Eddington. Furthermore, it may be directly tested, as IC10 X-1 is a high inclination system that eclipses on a $\sim$35 hour period \citep{prest07,sf08}; if our scenario is correct, then we would observe photo-electric absorption dips on the orbital period, and could determine the size of its corona from the ingress time.

\section*{Acknowlegments}
We thank the anonymous referee for their constructive and informative comments.  This work is based on observations with XMM-Newton, an
ESA science mission with instruments and contributions directly
funded by ESA member states and the US (NASA). Astronomy research at the Open University is funded by an STFC Rolling Grant.

\bibliographystyle{aa}
\bibliography{mnrasm31}
\label{lastpage}

\end{document}